\documentclass[runningheads]{llncs}
\usepackage{graphicx}
\usepackage{geometry}
\usepackage{booktabs}
\usepackage{float}

\begin{document}
\title{Evolving AI for Wellness: Dynamic and Personalized Real-time Loneliness Detection Using Passive Sensing\thanks{This publication has emanated from research conducted with the financial support of Science Foundation Ireland under Grant number 18/CRT/6222}}
\titlerunning{Evolving AI for Wellness}
%
\author{Malik Muhammad Qirtas\inst{1}\orcidID{0000-0001-7644-161X} \and
Evi Zafeiridi\inst{2}\orcidID{0000-0001-7986-5442} \and
Eleanor Bantry White\inst{3}\orcidID{0000-0002-7663-6836} \and
Dirk Pesch\inst{4}\orcidID{0000-0001-9706-5705}}
%
%
\institute{School of Computer Science and Information Technology, University College Cork, Ireland
\email{malik.qirtas@cs.ucc.ie}\\
 \and
 School of Computer Science and Information Technology, University College Cork, Ireland\\
\email{EZafeiridi@ucc.ie}
 \and
School of Applied Social Studies, University College Cork, Ireland\\
\email{E.BantryWhite@ucc.ie}
\and
School of Computer Science and Information Technology, University College Cork, Ireland\\
\email{dirk.pesch@ucc.ie}}
\maketitle              
\begin{abstract}
Loneliness is a growing health concern as it can lead to depression and other associated mental health problems for people who experience feelings of loneliness over prolonged periods of time. Utilizing passive sensing methods that use smartphone and wearable sensor data to capture daily behavioural patterns offers a promising approach for the early detection of loneliness. Given the subjective nature of loneliness and people's varying daily routines, past detection approaches using machine learning models often face challenges with effectively detecting loneliness. This paper proposes a methodologically novel approach, particularly developing a loneliness detection system that evolves over time, adapts to new data, and provides real-time detection. Our study utilized the Globem dataset, a comprehensive collection of passive sensing data acquired over 10 weeks from university students. The base of our approach is the continuous identification and refinement of similar behavioural groups among students using an incremental clustering method. As we add new data, the model improves based on changing behavioural patterns. Parallel to this, we create and update classification models to detect loneliness among the evolving behavioural groups of students. When unique behavioural patterns are observed among student data, specialized classification models have been created. For predictions of loneliness, a collaborative effort between the generalized and specialized models is employed, treating each prediction as a vote. This study's findings reveal that group-based loneliness detection models exhibit superior performance compared to generic models, underscoring the necessity for more personalized approaches tailored to specific behavioural patterns. These results pave the way for future research, emphasizing the development of finely-tuned, individualized mental health interventions.

\keywords{Grouping \and Loneliness \and Mobile sensing \and Passive sensing \and Smartphone}
\end{abstract}

\section{Introduction}
Loneliness is a growing global issue, with many people reporting it as a primary source of their unhappiness \cite{cacioppo2008loneliness}. It is an experience in which a person perceives a lack of quality social relationships \cite{peplau1985preventing}. Loneliness can lead to a variety of health problems, including difficulty sleeping, increased anxiety, persistent sadness, and a reduced immune response \cite{quadt2020brain,qirtas2023relationship,qirtas2022detecting,lim2020understanding}. While many people feel lonely at times, it becomes a concerning issue when it lasts for an extended period of time \cite{campbell2013loneliness}. Loneliness and mental health are closely linked. People with mental health issues are over twice as likely to feel lonely compared to those without such issues \cite{beutel2017loneliness}. Given the widespread rise of loneliness following the COVID-19 pandemic and its negative effects, it is crucial to detect loneliness early to reduce its potential harm.

The increasing use of smartphones and wearables has opened up new avenues for continuous and unobtrusive behavioural monitoring of individuals. Their ubiquitous nature and built-in sensors have made them effective tools for tracking user behaviours and their daily routines \cite{cornet2018systematic,sheikh2021wearable}. Passive sensing, an innovative technique where smartphones and wearables collect data without a user's active involvement, is leading this transformation. This mode of continuous data collection offers a more comprehensive view of an individual's behavioural patterns over time. When combined with established clinical scales for mental health detection, these digital footprints can be processed into biomarkers of mental health \cite{qirtas2022loneliness}. Specifically, by analyzing these digital biomarkers, researchers can extract signs of loneliness in real-time. As an unintrusive and economical method, passive sensing holds immense promise for revolutionizing the early detection of conditions such as loneliness, depression, or anxiety, opening  avenues for more tailored and personalized interventions.

In recent research, passive sensing methods have been employed to collect data, which is then analyzed using machine learning techniques to detect loneliness \cite{pulekar2016autonomously,doryab2019identifying,sarhaddi2022predicting,jafarlou2023objective,hayes2007distributed,qirtas2022privacy}. However, a common limitation of these studies is their reliance on generalized models that process all available data collectively. This approach, while effective in identifying broad behavioural patterns, often fails to account for the significant variations in daily living patterns across different individuals, potentially impacting the performance of loneliness detection models. Moreover, these systems have an emphasis on analysis that occurs after data collection. This limitation hinders their capacity to offer timely intervention and early detection in real time. This gap underscores the need for a more nuanced approach that not only acknowledges individual behavioural differences by identifying distinct sub-groups but also incorporates a mechanism for real-time monitoring. Our proposed methodology addresses these limitations by offering a tailored solution that adapts to individual behavioural nuances and provides real-time detection insights, setting a new direction in loneliness detection using passive sensing technologies. Our study leverages passive sensing data gathered from smartphones and wearables, which continuously monitor various aspects of the participating students' daily lives. This data encompasses a range of behavioural indicators such as physical activity patterns, social interactions inferred from call and message logs, and digital footprints like app usage, step count, sleep patterns and location tracking. These parameters offer a holistic view of a student's routine and social engagement, key factors influencing loneliness.

The basic idea of this paper is based on our earlier work \cite{qirtas2023personalising} but with some major modifications. At the core of our methodology are two intertwined components: Incremental Clustering and Incremental Classification. The clustering will be used to detect subgroups of students with the same behavioural patterns. This will continuously integrate new data into the existing model, refining behavioural groups over time. The incremental classification part will identify lonely individuals within these continually refining student groups. In parallel, we will be updating our main classification model. When new behavioural patterns appear, we will create a specific specialised model for them. For predicting loneliness, we will be using a Multi-Model Voting system that consider predictions from both the main and specialized models. Each prediction is like a 'vote'. The final decision will be based on either most votes or a system that weighs the trustworthiness of each model. The novelty of our approach lies in its dynamic and incremental nature, which addresses a gap in existing loneliness detection methodologies. Unlike static models, our system continually evolves, integrating new data to refine and update its understanding of student behaviours and loneliness indicators. This adaptability is crucial in capturing the transient and fluctuating nature of student lifestyles. This fusion of behavioural grouping and continuous loneliness detection among those groups, coupled with a multi-model approach, represents a significant methodological advancement in real-time loneliness detection.

\section{Methodology}

\begin{table}[H]
\centering
\caption{Study Information and Participant Demographics. Gender acronyms: F: Female, M: Male, NB: Non-binary. Racial acronyms: A: Asian, B: Black or African American, H: Hispanic, N: American Indian/Alaska Native, PI: Pacific Islander, W: White, NA: Did not report. The \& symbol denotes participants who identified with multiple races.}
\label{tab:demographicsTable}
\begin{tabular}{l|l}
\toprule
\textbf{Category} & \textbf{Data} \\
\midrule
Participants & Total: 218 \\
 & Gender: F 111, M 107 \\
 & Ethnicity: A 102, B 6, H 10, N 2, PI 1, W 70, \\
 & A\&B 1, A\&W 16, H\&W 2, B\&W 2, \\
 & A\&H\&W 1, B\&H\&W 1, H\&N\&W 1, NA 3 \\
Ground Truth & Pre-study 10-items UCLA scale \\
 & Post-study 10-items UCLA scale \\
Sensor Data & Bluetooth, WIFI, Call Logs, Location, Phone Usage, Physical Activity, Sleep \\
\bottomrule
\end{tabular}
\end{table}

\subsection{Dataset}
We have worked with a dataset from University of Washington, collected during the Spring quarter of 2019 over 10 weeks from March to June \cite{xu2022globem}. This period was chosen to consider any seasonal influences. The dataset, named DS-2, includes information from 218 undergraduate students, majority females, who joined through email and social media invitations. Please refer to Table~\ref{tab:demographicsTable} for detailed information about participants' demographics.

For the data collection, the AWARE smartphone application, available for both iOS and Android platforms, was employed. This application operates continuously in the background without requiring active user interaction \cite{ferreira2015aware}. Additionally, a Fitbit was used to gather sleep and physical activity data. To offer an in-depth understanding of these patterns, the study integrated readings from various 'sensors' in both smartphone and Fitbit, including Bluetooth, WiFi, location, accelerometer, gyroscope, microphone, and light sensors. The data collection initiative received ethical approval from the University of Washington's IRB (IRB number:STUDY00003244), and every participant gave informed consent. Adhering strictly to anonymization protocols ensured data confidentiality, with direct identifiers limited to the core data team. Moreover, any data from participants choosing to withdraw was immediately removed from the database.

The study used the revised 10-item UCLA loneliness scale to assess loneliness \cite{knight1988some} and to label the sensor data. Participants completed this questionnaire at the beginning and end of the study, rating questions on a scale of 1 ("never") to 4 ("always"). After reverse scoring five items, the scores were combined, resulting in a total score ranging between 10 to 40. Higher scores suggest increased feelings of loneliness. We used these scores as the ground truth for experiences of loneliness, with scores above 20 indicating higher feelings of loneliness.

\begin{figure}[H]
\centering
\includegraphics[width=\textwidth, height=7.5cm, keepaspectratio]{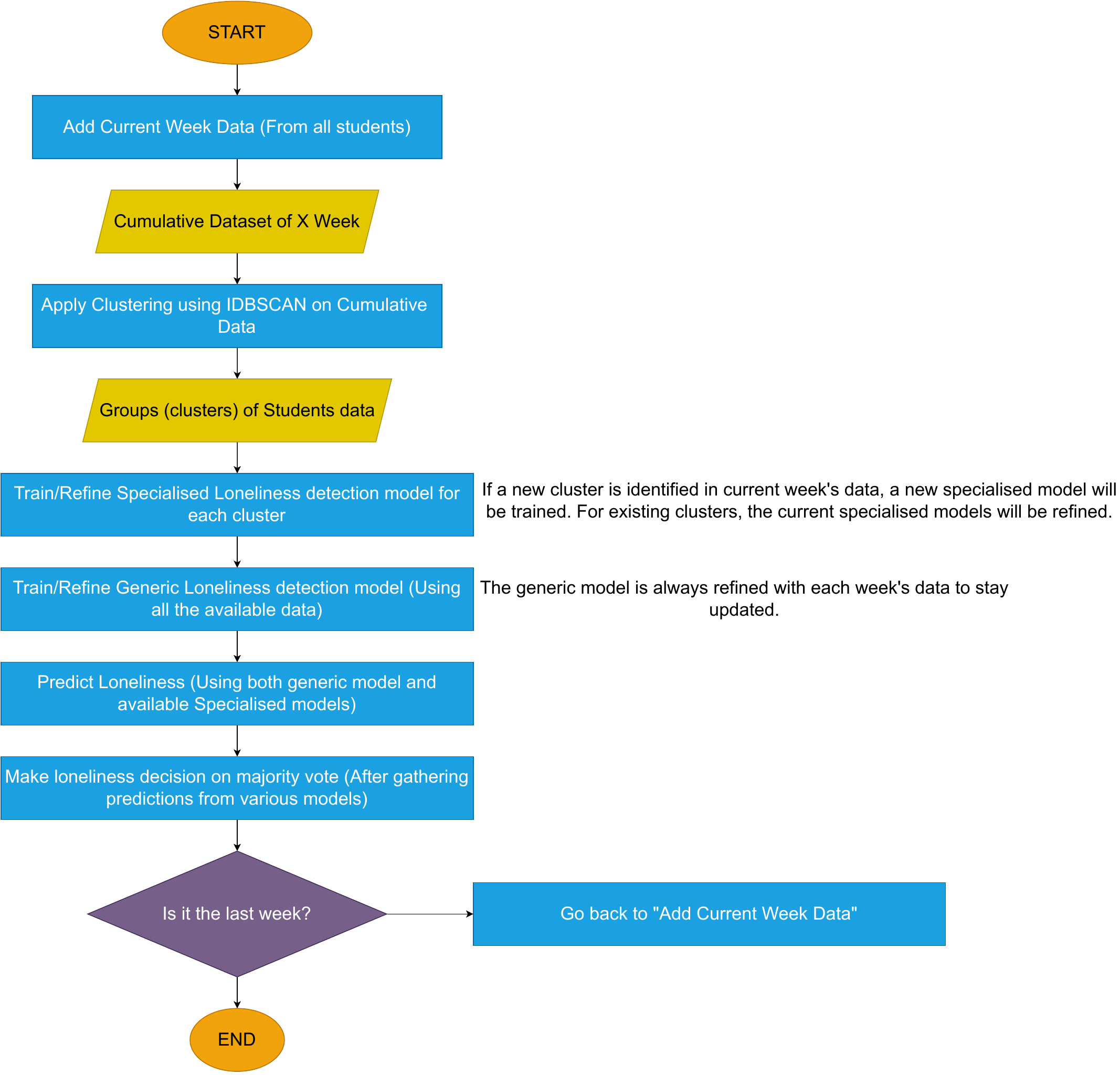}
\caption{Flowchart depicting the iterative process of incremental clustering, model refinement, and loneliness prediction across 10 weeks.}
\label{fig:flowchart}
\end{figure}

\subsection{Data Preprocessing}
During preprocessing, we refined the dataset to include only the data from 205 out of 218 students who completed the post-study loneliness questionnaire. We took loneliness detection as a binary classification problem by dividing into two main categories based on loneliness scores: students with higher feelings of loneliness (scoring more than 20 on the UCLA loneliness scale) and those with below the score of 20. Of the 205 students, 87 were classified as 'lonely', and the other 118 were not, giving us a distinct class distribution for our analysis. To derive comprehensive behavioural insights from raw sensor data, each data stream was segmented into daily intervals, spanning from 12:00 am to 11:59 pm. These intervals were further subdivided into distinct time segments: morning (6 am - 12 pm), afternoon (12 pm - 6 pm), evening (6 pm - 12 am), and night (12 am - 6 am). The division into specific time segments serves an important role in capturing nuanced behavioural patterns, as individuals tend to engage in distinct activities during different parts of the day. To address the issue of missing values, we initially removed all records with outliers. Subsequently, we filled in missing data for each student using the median value for continuous data features specific to a session. For categorical data, we used the mode value of a particular feature for that session. The dataset, processed using the Reproducible Analysis Pipeline for Data Streams (RAPIDS) \cite{vega2020rapids}, produced digital biomarkers by quantifying students' behavioural patterns on a per-participant and per-session basis. These patterns, like routines and variability, were measured using metrics such as counts, standard deviations, and entropy. The detailed overview of these features are provided in RAPIDS documentation \cite{vega2020rapids,knuthwebsite}. Features were then grouped based on sessions (day, evening, night). Categorical features underwent one-hot encoding to achieve integer representation. Numerical data were normalized using min-max scaling, which adjusted each value to a range between 0 and 1. Given that clustering is more effective with a reduced feature set, we applied principal component analysis (PCA) to the entire dataset for dimensionality reduction.

\subsection{Evolving Student Behavioural Profiles via Incremental Clustering}

\begin{figure}[H]
  \centering
  \begin{minipage}[c]{0.45\textwidth}
    \includegraphics[width=\textwidth]{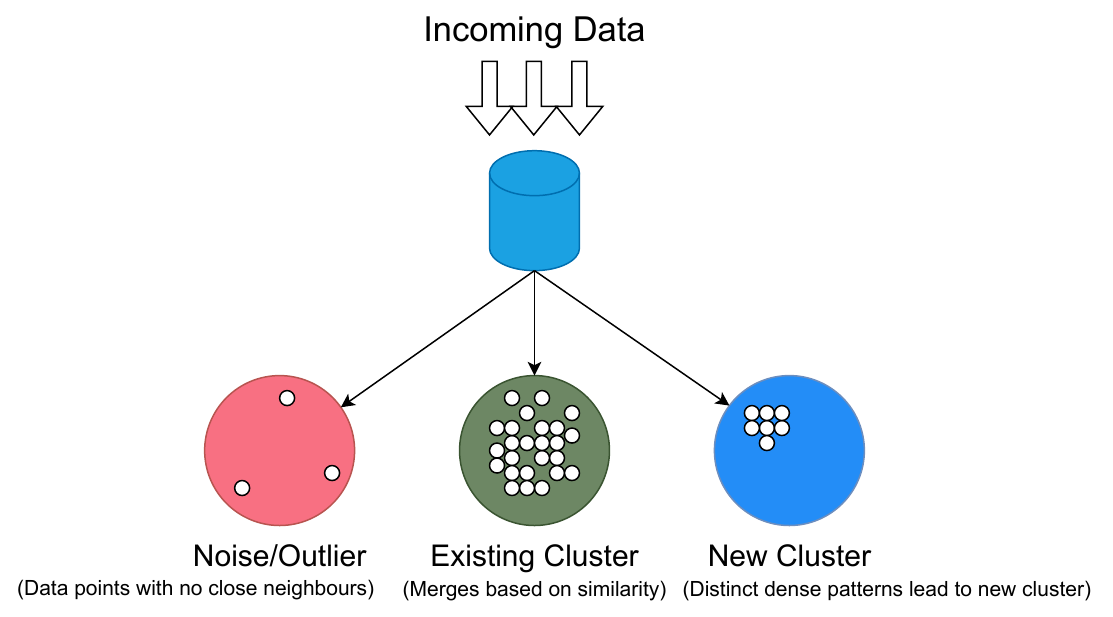}
    \caption{Visualization of Real-time Data Clustering: Segregation into Noise, Existing, and Emergent Clusters}
    \label{fig:clustering}
  \end{minipage}
  \hfill 
  \begin{minipage}[c]{0.45\textwidth}
    \includegraphics[width=\textwidth]{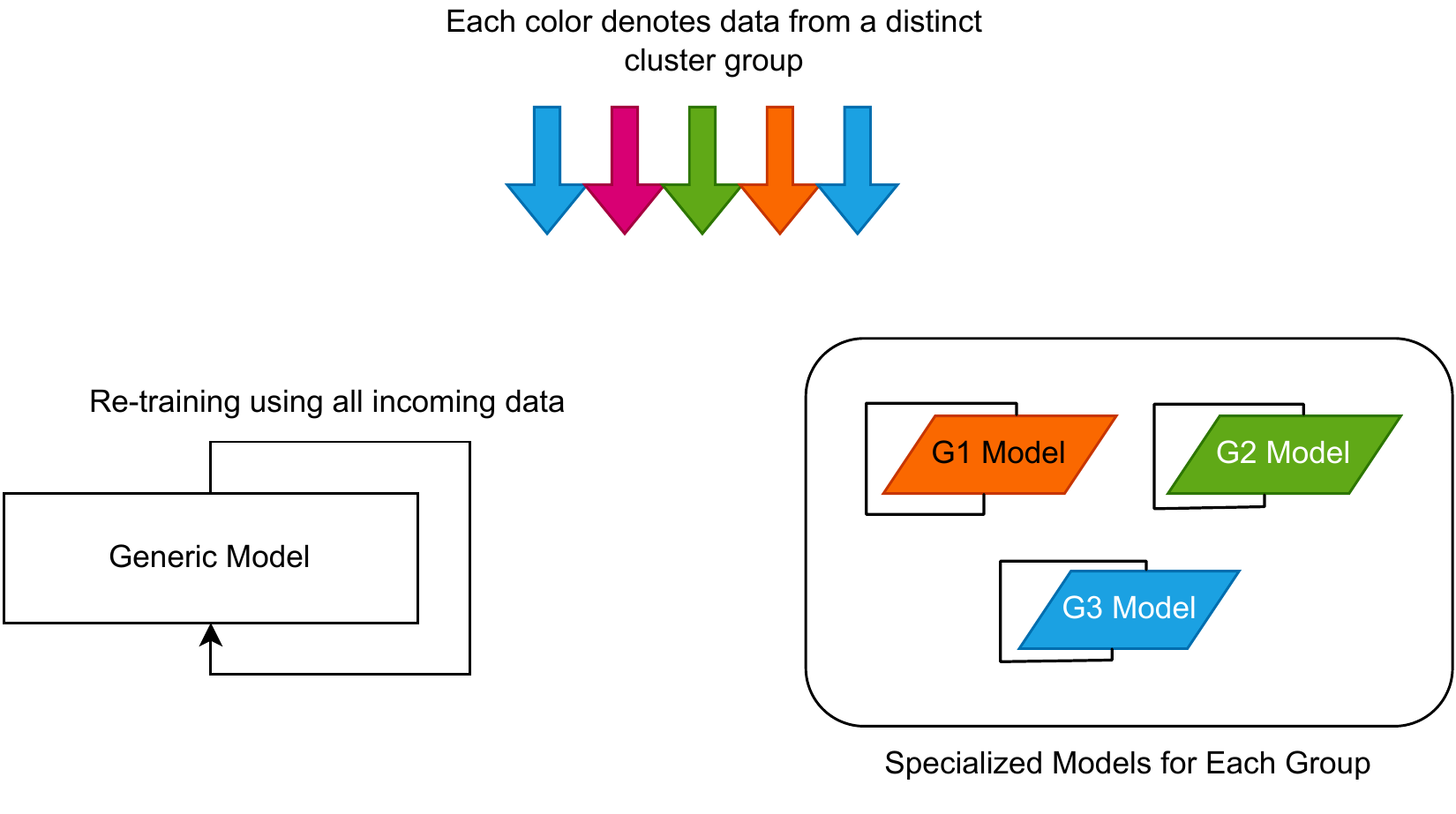}
    \caption{Personalized Loneliness Detection Using behavioural Groups}
    \label{fig:classification}
  \end{minipage}
\end{figure}

Incremental clustering is a pivotal aspect of our approach, primarily because student behaviours always change. This approach not only adapts to the constantly evolving nature of student behaviours but also eliminates the need for model retraining from scratch on the arrival of new data. The incremental clustering method we use is a dynamic process that incorporates real-time data into our evolving behavioural model, as seen in the Fig.\ref{fig:clustering}. Upon the arrival of new data, it is compared to pre-existing clusters. Every new piece of data is evaluated for its fit: it may either seamlessly integrate into an existing cluster if it aligns with the established pattern of the group, or it can be identified as an outlier, an isolated data point that does not match any known group. If the new data shows a distinct and tightly grouped pattern, it acts as the origin of a new cluster. Our flexible and progressive learning approach ensures that our model stays flexible and up-to-date, effectively reflecting the constantly changing behaviours of the students.

Before the clustering process, a crucial step involves standardizing the dataset to maintain consistency. This means setting the value of each feature to a standard range. This ensures all features have the same effect on distance calculations for clustering. This kind of uniformity is very important because it keeps any one feature from having an unfair effect on the clustering because of differences in scale. This results in a more accurate and representative clustering outcome. After preparing the data, we proceed to the clustering phase. Among the many available clustering algorithms, we selected Incremental Density-Based Spatial Clustering of Applications with Noise (IDBSCAN) \cite{goyal2009efficient} because this can be updated in an incremental fashion and it has shown good performance in our earlier work \cite{qirtas2023personalising}. Essentially, IDBSCAN is an enhanced version of the conventional DBSCAN algorithm that has been intelligently modified to process data in an incremental way. This is seen in its ability to provide a hierarchical depiction of data density distribution. This approach enables it to effectively incorporate new data into existing clusters or identify them as anomalies. At times, when there are new data points that have a clear difference in density, IDBSCAN is able to detect the formation of new clusters. Upon obtaining new data, the algorithm assesses it in relation to the existing clusters. This evaluation utilizes proximity and density criteria to ascertain the most suitable match for each new data point. When the 'eps' parameter is set to 0.5, and the minimum density threshold is 0.1, IDBSCAN evaluates whether to include a data point in an existing cluster, classify it as noise, or mark it as the starting point of a new cluster. 

\subsection{Real-Time Loneliness Detection through Incremental Classification}

Our primary objective with incremental classification is to detect loneliness in behavioural subgroups in real time. These subgroups emerge following the phase of clustering. Each subgroup consists of students with common behavioural patterns. Given the dynamic nature of changing student behaviours, a static classification method may rapidly become ineffective or imprecise. As a result, our strategy supports continuous real-time loneliness detection. Initially, we approach loneliness detection as a binary classification problem, marking individuals as "lonely" (1) or "not lonely" (0) based on their UCLA survey responses. Specifically, students with scores above 20 are categorized as lonely. Starting with this initial stage, we develop a foundational 'generic' model that is trained using initial data.

However, a single model may not always cover the nuanced differences revealed by distinct behavioural subgroups. To cater to the distinct features inherent in each student group, we introduce 'specialized' classification models. Each of these models is dedicated to a specific group and was trained using only students data from that group. By focusing on a smaller group, these models can better detect the nuances of that group, allowing them to make more personalized and precise predictions. Fig.\ref{fig:classification} shows the process of refining the data that leads to the generic model and then to models G1, G2, and G3 which are each specifically made for a different group.

To resolve the class imbalance in the training dataset, we employed the Synthetic Minority Oversampling Technique (SMOTE) \cite{chawla2002smote}, which generates synthetic data for under-represented classes, thereby assuring a more balanced training dataset for our classifiers. The application of SMOTE reduces biases towards the majority class, thereby enhancing the performance of classification models \cite{elreedy2019comprehensive}. We developed loneliness prediction models using logistic regression, random forest, support vector machine, and XGBoost algorithms. Rather than taking a one-size-fits-all approach, these models were developed specifically for each behavioural group, with a focus on personalized predictions. The performance of our models was evaluated using metrics such as accuracy, precision, recall, and the F1 score. While accuracy measures the overall reliability of a prediction, precision and recall provide insight into truly positive predictions. The F1 score integrates precision and recall to provide a comprehensive evaluation of a model's performance. To assure an accurate evaluation, we used the k-fold cross-validation technique with k set to 10. By dividing the data into ten sections and cyclically using each as a test subset, we ensure that each data point is evaluated at least once, allowing for a more accurate estimation of the model's performance on unseen data.

\subsection{Multi-Model Voting}

Our approach is built around the use of Multi-Model Voting. This method is based on the idea that insights from multiple models lead to more reliable and consistent outcomes. It aims to use the combined expertise from multiple models. The idea behind this approach comes from the fact that different models may have different strengths and weaknesses. When we use more than one model, we look at data from various perspectives, reducing the risks that come with depending on one point of view. That can be particularly helpful when the data is not skewed or is full of errors, because a single model might fail or produce imprecise results in such cases.

Our approach begins by collecting predictions from a number of models. This includes not just a generic model but also a set of specialized models, each tailored to specific behavioural groups.  The diversity of these predictions provides a complex collection of views, with each model providing its own interpretation based on its understanding and area of specialty. After collecting predictions from multiple models, our goal is to combine these views into a single decision. The majority rule is our main guide, meaning we choose the outcome that most models agree on. This approach helps us achieve a clear and trustworthy classification.

\begin{table}[H]
\centering
\begin{tabular}{|p{1.5cm}|p{1cm}|p{9.1cm}|p{1.9cm}|}
\hline
\textbf{Week Interval} & \textbf{Group} & \textbf{behavioural Characteristics} & \textbf{Loneliness Score Range} \\
\hline
Weeks 1-4     & G1    & High physical activity, Frequent calls \& SMS, High unique Bluetooth encounters, Regular location changes, Moderate phone usage, Average sleep duration & 10-18  \\
\hline
              & G2    & High physical activity, Average calls \& SMS, Average unique Bluetooth encounters, Moderate phone usage, Average sleep duration & 15-26       \\
\hline
              & G3    & Low physical activity, Fewer calls \& SMS, Fewer unique Bluetooth encounters, Moderate location changes, High phone usage, Short sleep duration & 16-30   \\
\hline
Weeks 5-7     & G1    & High physical activity, Frequent calls \& SMS, High unique Bluetooth encounters, Regular location changes, Moderate phone usage, Average sleep duration & 10-22   \\
\hline
              & G2    & Reduced physical activity, Fewer calls \& SMS, Average unique Bluetooth encounters, Moderate location changes, Increased phone usage, Reduced sleep duration & 15-32   \\
\hline
              & G3    & Low physical activity, Fewer calls, Average unique Bluetooth encounters, Fewer location changes, High phone usage, longer Short duration & 18-36    \\
\hline
              & G4    & Moderate physical activity, Variable calls \& SMS, High unique Bluetooth encounters, Regular location changes, Moderate phone usage, Variable sleep duration & 14-28    \\
\hline
Week 8        & G1    & High physical activity, Frequent calls \& SMS, High unique Bluetooth encounters, Regular location changes, Moderate phone usage, Average sleep duration & 10-20        \\
\hline
              & G2    & High physical activity, Average calls \& SMS, Average unique Bluetooth encounters, Regular phone usage, Average sleep duration & 15-25          \\
\hline
              & G3    & Low physical activity, Variable calls \& SMS, Fewer unique Bluetooth encounters, Moderate location changes, High phone usage, Short sleep duration & 18-36        \\
\hline
Weeks 9-10  & G1    & High physical activity, Frequent calls \& SMS, High unique Bluetooth encounters, Regular location changes, Moderate phone usage, Average sleep duration & 10-20         \\
\hline
              & G2    & High physical activity, Average calls \& SMS, Average unique Bluetooth encounters, Regular phone usage, Average sleep duration & 16-32     \\
\hline
              & G3    & Low physical activity, Variable calls \& SMS, Fewer unique Bluetooth encounters, Moderate location changes, High phone usage, Short sleep duration & 18-36     \\
\hline
              & G4    & Low physical activity, Variable calls \& SMS, Variable unique Bluetooth encounters, Moderate location changes, Moderate phone usage, Average sleep duration            & 14-35     \\
\hline
\end{tabular}
\caption{Summary of behavioural Patterns and Associated Loneliness Scores by Group Over 10 Weeks}
\label{table:behavioural_profiles_loneliness}
\end{table}

\begin{figure}[H]
  \centering
  \includegraphics[width=0.9\textwidth]{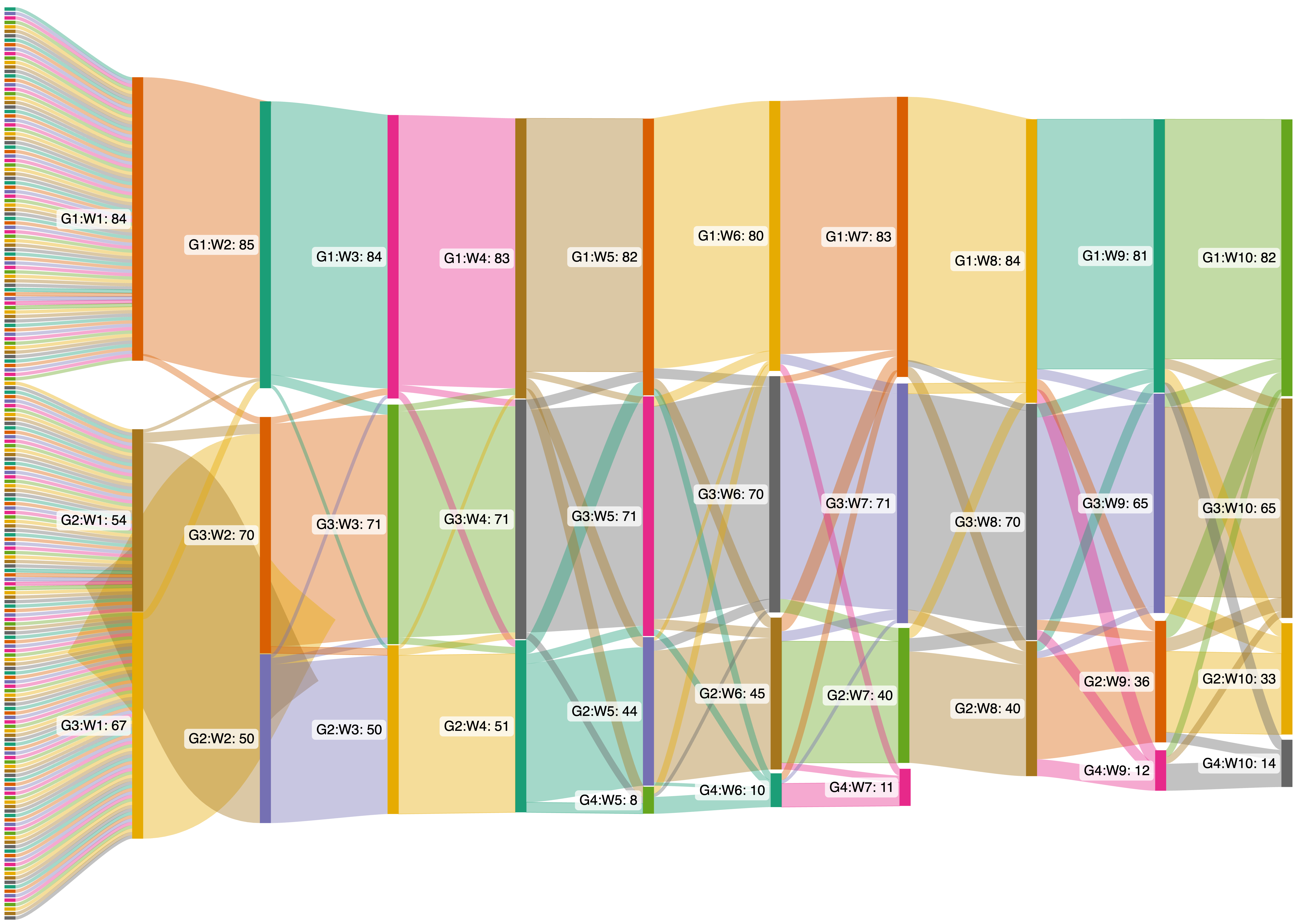}
  \caption{Visualization of Students Temporal behavioural Group Dynamics over a 10-Week Period (G1:W1:84 denotes Group 1 in Week 1 with 84 students)}
  \label{fig:dynamics}
\end{figure}

\section{Results}

\subsection{Temporal Evolution of Student Groups and Behavioural Profiles}
Our analysis utilizing clustering revealed dynamic shifts in student behavioural groups over a 10-week study period, as visually represented in the Fig. \ref{fig:dynamics}. Initially, three groups G1, G2, and G3 were identified, characterized by varying levels of physical activity, phone usage, sleeping patterns and social interaction. G1, the largest group, was marked by high physical and social activity, G2 by a moderate level, and G3 by the lowest activity. These groups displayed a relatively stable pattern of behaviour, with minor fluctuations in size indicating some movements of students among the groups. The emergence of a distinct group, G4, during week five indicated a change from the initial patterns of behaviour. This group was identified by a mix of moderate physical activity and moderate phone usage. This period marks a shift from the stability of the early weeks to a more dynamic phase where students began to explore different behavioural patterns, as evidenced by the formation of this new group.

To analyze the behavioural patterns of each group, we calculated the weekly average of their features. This approach provides a simple and straightforward way to understand the behavioural tendencies of the groups, which allows for additional analysis and interpretation of the results. For example, G1 showed frequent location changes and numerous Bluetooth encounters. In contrast, G3's lower activity was indicated by fewer Bluetooth encounters and more stationary location data. Corresponding with these behavioural profiles, loneliness scores derived from the UCLA scale were assigned to each group. G1's profile with high activity corresponded with lower loneliness scores, whereas G3's profile with lower levels of activity and social engagement showed with higher loneliness scores. G2 and G4 exhibited scores that reflected their intermediate and variable engagement levels, respectively. Throughout the study, groups underwent transitions, with a notable shift in G2's size diminishing and G4's increasing, particularly in the final weeks. Table~\ref{table:behavioural_profiles_loneliness} summarizes these observations, providing detailed information on the behavioural tendencies and associated loneliness scores for each group. The scores are presented to illustrate potential trends across the groups during the study period. As the term advanced, notable shifts were observed: the reabsorption of students from G4 back into the original three groups in week 8, and the re-emergence of G4 in weeks 9 and 10 with an increased student count.

\subsection{Loneliness Detection Using Generic Models}
\begin{figure}[H]
    \centering
    \begin{minipage}{.42\textwidth}
        \includegraphics[width=\linewidth]{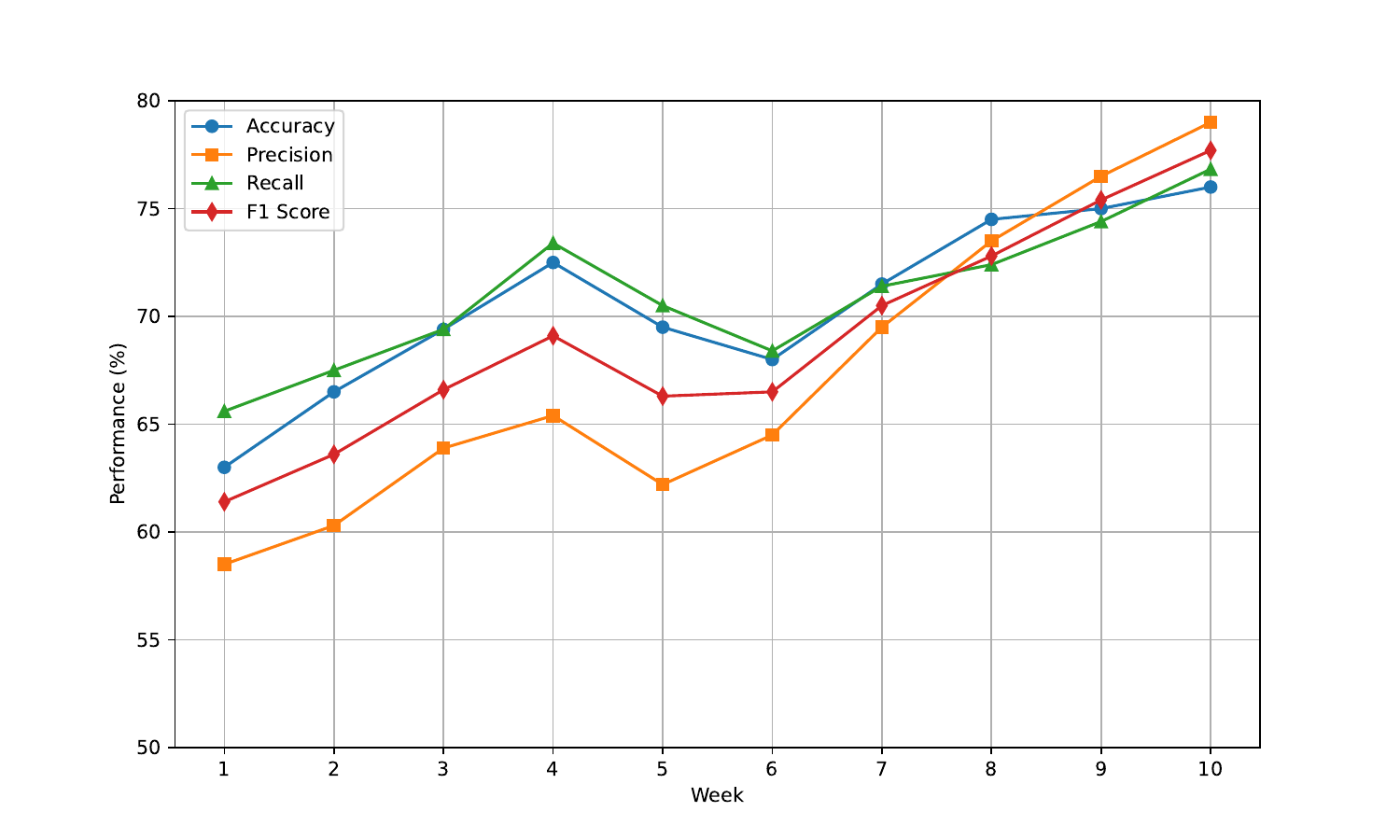}
        \caption{Temporal Performance of the XGBoost Model for Loneliness Detection Across a 10-Week Period}
        \label{fig:generic_XGBoost}
    \end{minipage}%
    \hfill
    \begin{minipage}{.42\textwidth}
        \includegraphics[width=\linewidth]{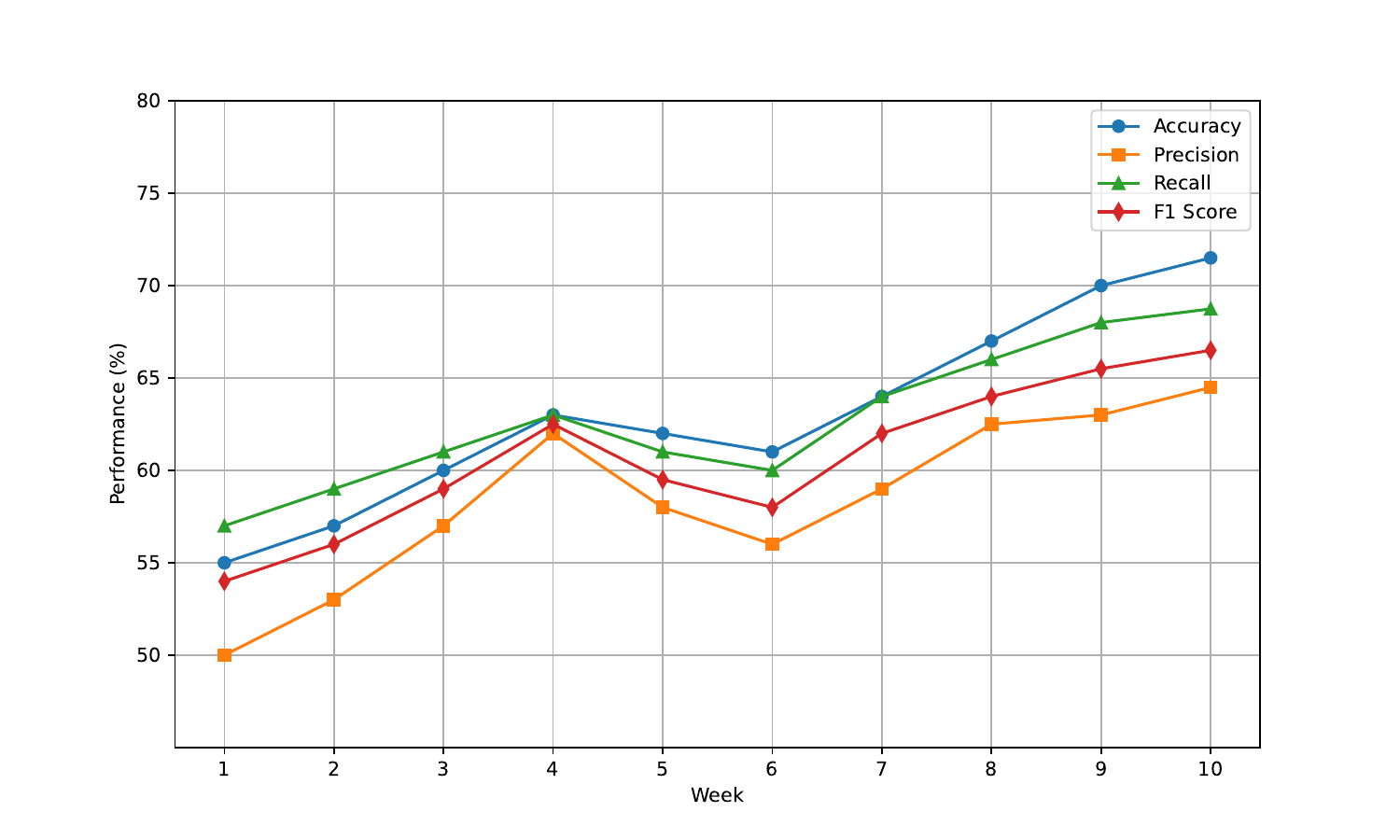}
        \caption{Temporal Performance of the SVM Model for Loneliness Detection Across a 10-Week Period}
        \label{fig:generic_SVM}
    \end{minipage}
        \hfill
    \begin{minipage}{.42\textwidth}
        \includegraphics[width=\linewidth]{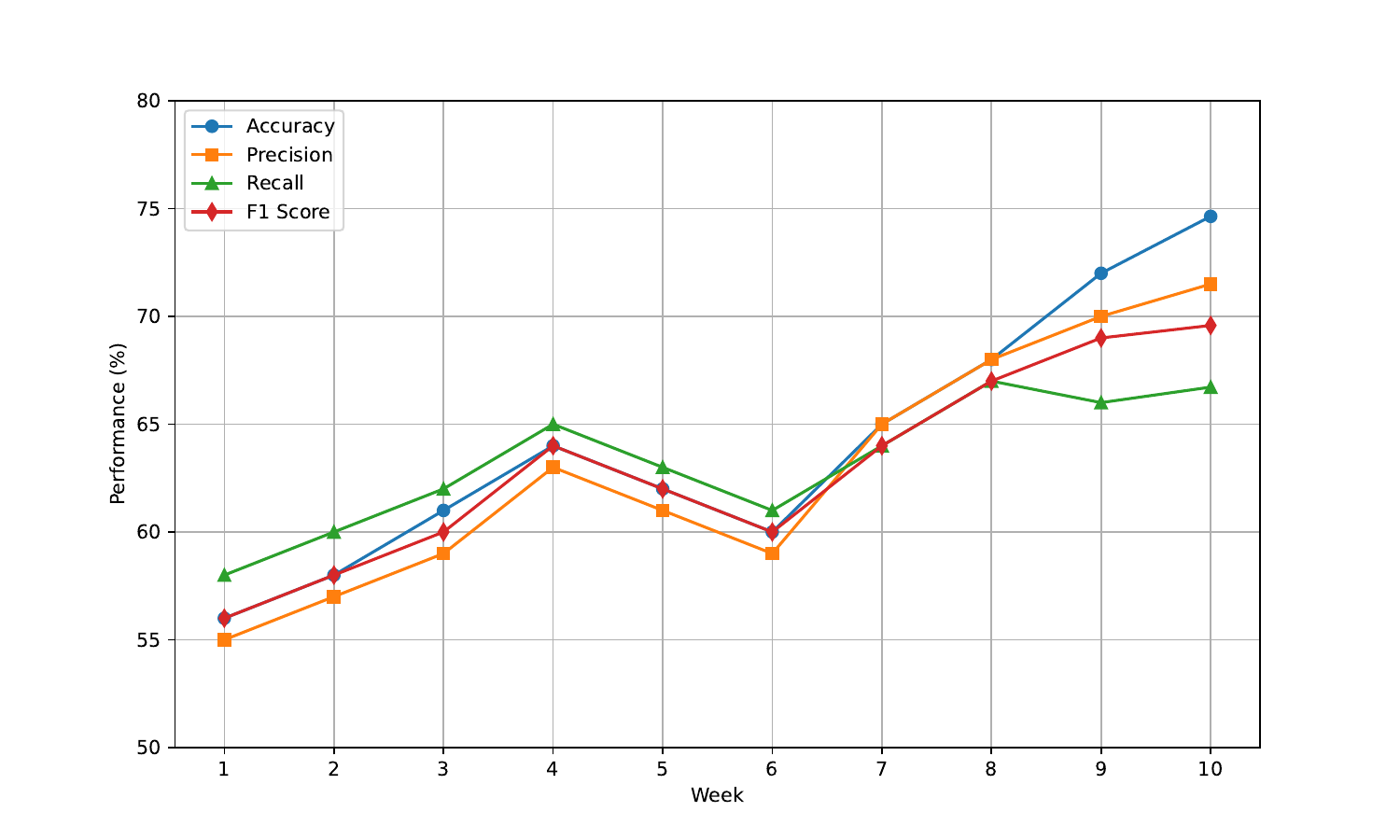}
        \caption{Temporal Performance of the Random Forest Model for Loneliness Detection Across a 10-Week Period}
        \label{fig:generic_RandomForest}
    \end{minipage}
            \hfill
    \begin{minipage}{.42\textwidth}
        \includegraphics[width=\linewidth]{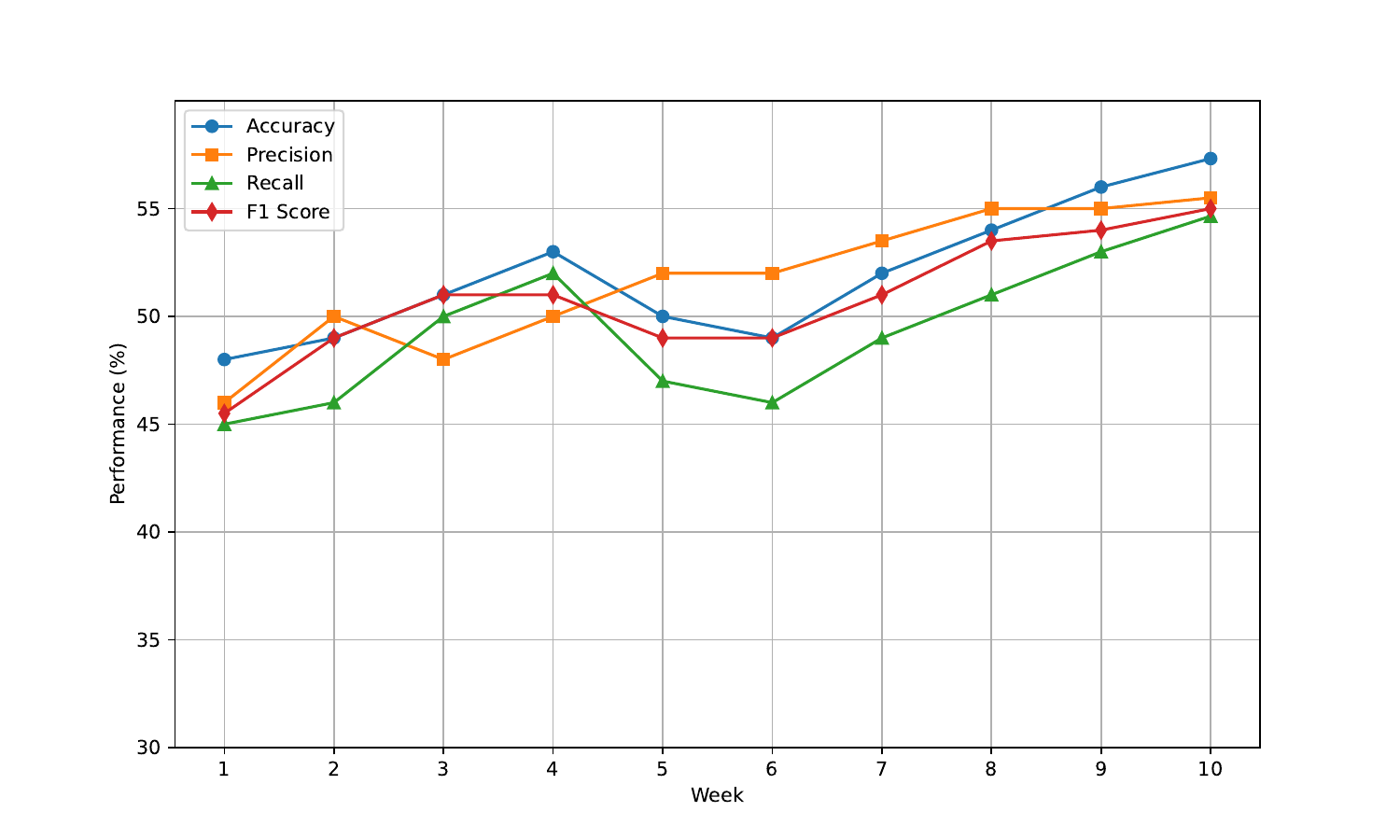}
        \caption{Temporal Performance of the Logistic Regression Model for Loneliness Detection Across a 10-Week Period}
        \label{fig:generic_LogistcRegression}
    \end{minipage}
\end{figure}

The performance of the generic machine learning models, which underwent weekly incremental training on the dataset, is presented in Table~\ref{tab:generic_model_performance}. This table has the outcomes of four binary classification algorithms: XGBoost, Support Vector Machine (SVM), Random Forest, and Logistic Regression. Each model's adaptability and refinement in detecting loneliness are reflected in the progressive improvement of accuracy, precision, recall, and F1 score metrics across the 10-week period. Fig~\ref{fig:generic_XGBoost} to Fig~\ref{fig:generic_LogistcRegression} shows the weekly performance of each classification model for loneliness detection. Although there were initial fluctuations, especially during the study's midpoint, the final week's results indicate that the models predictive capabilities improved as they processed an increasing quantity of data.

\begin{table}[H]
\centering
\caption{Performance of Generic Models for Loneliness Detection}
\label{tab:generic_model_performance}
\begin{tabular}{lcccc}
\toprule
\textbf{Model} & \textbf{Accuracy} & \textbf{Precision} & \textbf{Recall} & \textbf{F1 Score} \\
\midrule
XGBoost & 76.52\% & 79.00\% & 76.83\% & 77.5\% \\
Random Forest & 74.64\% & 71.50\% & 66.72\% & 69.58\% \\
SVM & 71.50\% & 64.49\% & 77.74\% & 68.50\% \\
Logistic Regression & 57.32\% & 43.50\% & 54.65\% & 48.53\% \\
\bottomrule
\end{tabular}
\end{table}

During the course of the 10-week study period, each generic model showed a trajectory of performance in the prediction of loneliness. The XGBoost model showed steady improvement, starting at 63\% accuracy in week 1 and reaching 76.5\% by week 10, with corresponding increases in precision, recall, and F1 score to 79\%, 76.83\%, and 77.5\%, respectively. The model performance displayed a minor dip around weeks 5 and 6, which was subsequently improved. Random Forest accuracy improved from 59\% to 74.64\%, with precision peaking at 71.5\% and recall at 66.72\%. The F1 score for this model finalized at 69.58\%. Similar to other models, Random Forest encountered week-to-week variances, particularly around weeks 5 and 6. The SVM model's accuracy improved from an initial 55\% to 71.5\% accuracy, while precision and recall also rose to 64.49\% and 77.74\%, respectively. The F1 score for SVM reached 68.5\%, indicating a consistent upward trend despite fluctuations in the mid-weeks. The logistic regression model shows a performance increase over a 10-week period, with accuracy starting below 40\% and rising to over 57.32\%. While all metrics demonstrate an upward trend, the F1 score shows some variability but still indicates overall performance improvement.

\begin{table}[H]
  \centering
  \caption{Performance of behavioural group based machine learning models classifying loneliness.}
  \label{tab:table1}
    \begin{tabular}{cccccccccccccccccccc}
\textbf{Algo} & \multicolumn{4}{c}{\textbf{G1}} &  & \multicolumn{4}{c}{\textbf{G2}} &  &  \multicolumn{4}{c}{\textbf{G3}} &  &  \multicolumn{4}{c}{\textbf{G4}}  \\ 
\cline{2-5} \cline{7-10} \cline{12-15} \cline{17-20}
        &  \textbf{Acc.}   & \textbf{Prec.}  & \textbf{Rec.} & \textbf{F1.}                                             &             
	&  \textbf{Acc.}   & \textbf{Prec.}      & \textbf{Rec.} & \textbf{F1.}   
 &             
	&  \textbf{Acc.}   & \textbf{Prec.}      & \textbf{Rec.} & \textbf{F1.}
 &             
	&  \textbf{Acc.}   & \textbf{Prec.}      & \textbf{Rec.} & \textbf{F1.}
 \\    \hline
 \textbf{XGBoost}   &  78.64   & 79.84   & 81.5  & 80.5                                 &
    & 80.15   & 78.63   & 82.44  & 80.74 & 
    & 76.55   & 78.34 & 73.74 & 75.97 &
    & 41.38   & 39.66 & 44.38 & 42.10        \\ 
\textbf{RF}   &  78.35   & 77.74   & 74.44  & 75.73                                 &
    & 79.00   & 80.45 & 78.55 & 79.34 & 
    & 75.53   & 78.43 & 80.14 & 79.55 &
    & 62.42   & 63.4 & 65.33 & 64.00        \\     
 \textbf{SVM}   &  77.43   & 71.52   & 80.45  & 75.38                                 &
    & 76.49   & 75.55 & 79.32 & 77.43 & 
    & 72.31   & 76.52 & 78.33 & 77.18 &
    & 53.41   & 57.53 & 55.29 & 56.52        \\   
    \textbf{LR}   &  64.29   & 54.32   & 63.71  & 58.45                                 &
    & 58.29   & 55.62 & 58.41 & 56.29 & 
    & 52.00   & 44.42 & 55.29 & 48.73 &
    & 42.48   & 39.84 & 44.23 & 41.96        \\  
 
      \bottomrule
    \end{tabular}
\label{tab:groupbasedmodels}
\begin{tabbing}
Algo:Binary Classification Algorithm; G1: Group 1; G2:Group 2; G3:Group 3; G4:Group 4; Acc:Accuracy(\%); \\ Prec:Precision(\%); Rec:Recall(\%)  F1:F1 Score(\%); RF:Random Forest; SVM:Support Vector Machine; \\ LR:Logistic Regression
\end{tabbing}
\end{table} 

\subsection{Loneliness Detection Using Group Based Models}
In order for classification algorithms to detect loneliness within each behavioural group, we have trained 4 classification models; XGBoost, Random Forest, Support Vector Machine(SVM) and Logistic Regression. The efficacy of loneliness detection models for each of the four groups is illustrated in Table-\ref{tab:groupbasedmodels}.

The XGBoost algorithm showed a strong performance across all groups, with the highest accuracy in G1 (78.64\%) and the lowest in G4 (41.38\%). Random Forest (RF) and Support Vector Machine (SVM) models exhibited consistent performance, with RF slightly outperforming SVM in most groups. Notably, the Random Forest model demonstrated remarkable robustness in G4, achieving an accuracy of 62.42\%, a substantial improvement over XGBoost for this group. Logistic Regression (LR), while the least effective of the models, still provided valuable insights, especially in distinguishing patterns in G1 and G2, where it achieved accuracies of 64.29\% and 58.29\%, respectively. When compared with generic models' performance as presented in Table \ref{tab:generic_model_performance}, group-based models generally showed an improvement in detecting loneliness. This trend was observed across most metrics and models, suggesting that a tailored approach to individual groups can enhance model sensitivity to loneliness indicators.

\begin{figure}[H]
    \centering
    \begin{minipage}{.42\textwidth}
        \includegraphics[width=\linewidth]{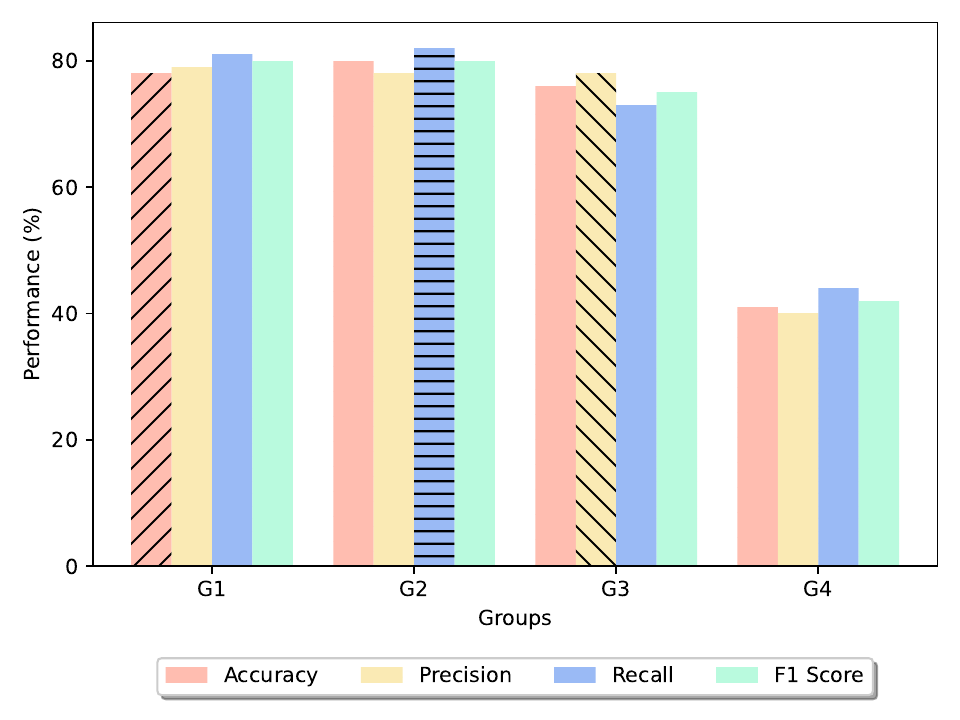}
        \caption{Performance of the Group based XGBoost Model for Loneliness Detection}
        \label{fig:graph1}
    \end{minipage}%
    \hfill
    \begin{minipage}{.42\textwidth}
        \includegraphics[width=\linewidth]{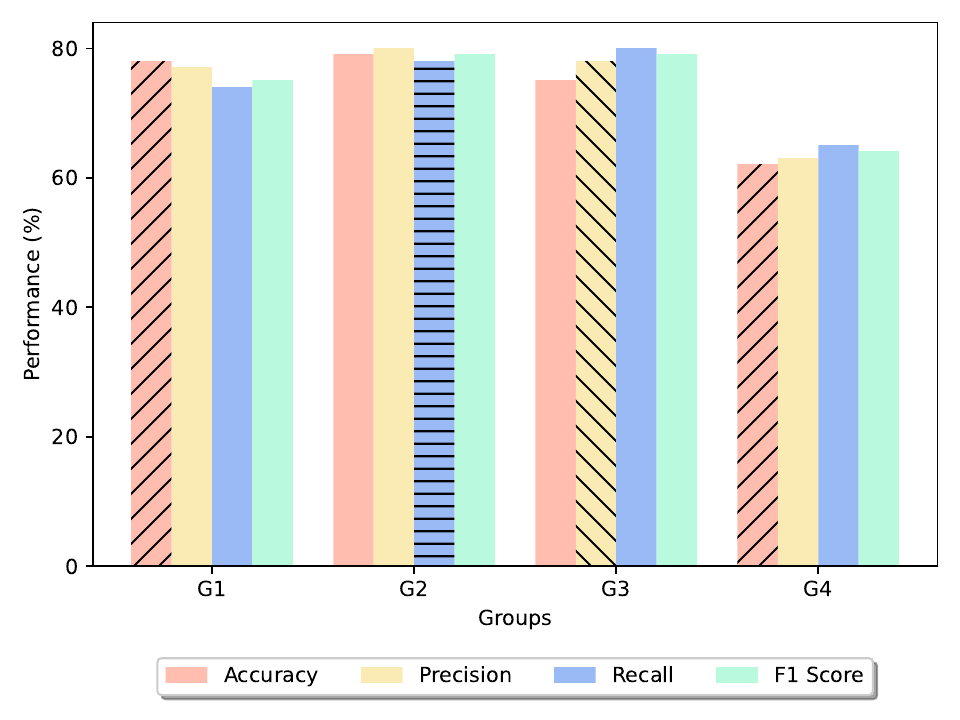}
        \caption{Performance of the Group based Random Forest Model for Loneliness Detection}
        \label{fig:graph2}
    \end{minipage}
        \hfill
    \begin{minipage}{.42\textwidth}
        \includegraphics[width=\linewidth]{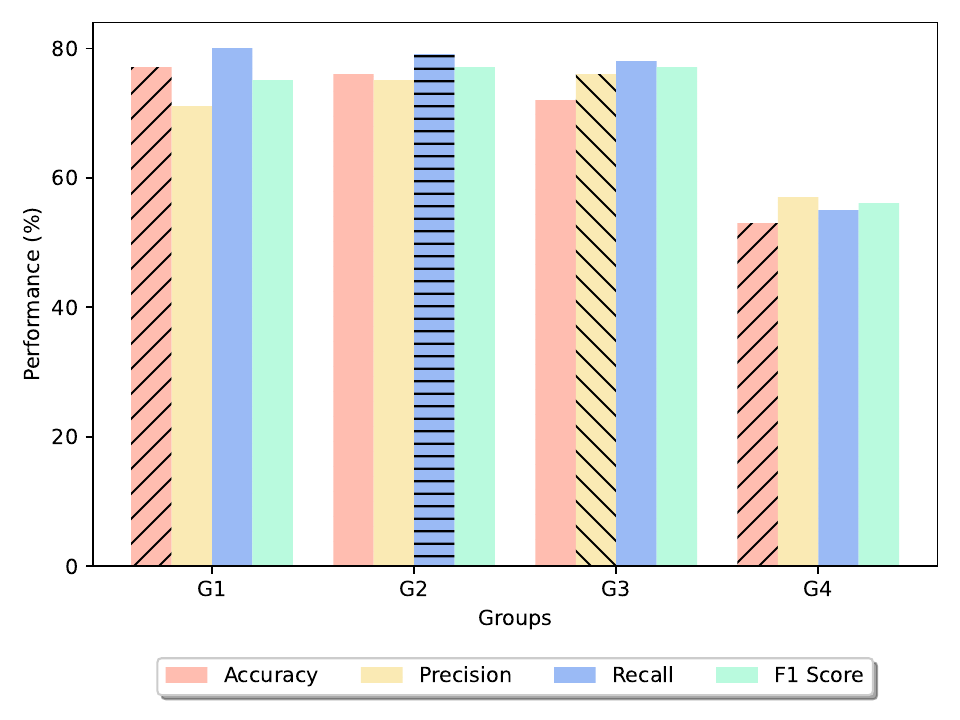}
        \caption{Performance of the Group based SVM Model for Loneliness Detection}
        \label{fig:graph2}
    \end{minipage}
            \hfill
    \begin{minipage}{.42\textwidth}
        \includegraphics[width=\linewidth]{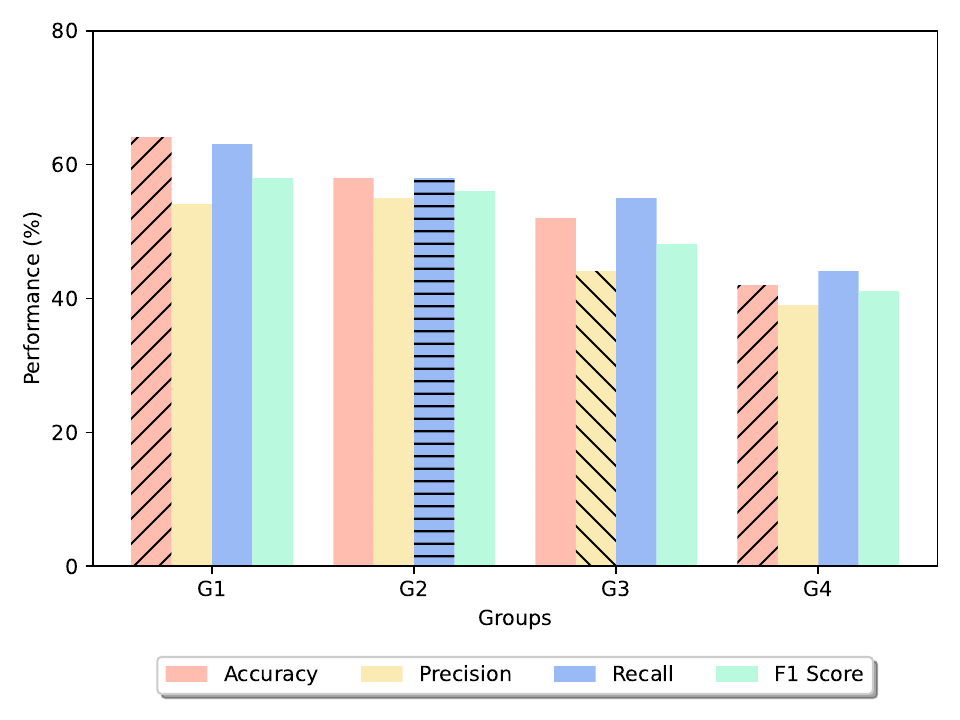}
        \caption{Performance of the Group based Logistic Regression Model for Loneliness Detection}
        \label{fig:graph2}
    \end{minipage}
\end{figure}

\subsection{Discussion}

This study presents a significant contribution to the field of loneliness detection among  students, employing a novel approach that combines passive sensing data from smartphones and wearables with machine learning techniques. The study highlights the effectiveness of group-based models, which performed better by targeting specific behavioural groups than generic models. The better performance is due to their ability to cut down on noise and variation in the dataset. This makes it possible to find more precise patterns of behaviour that are linked to loneliness. The results indicate that personalized models, including specific group dynamics can provide more precise estimations of loneliness than generic conventional methods. The results showed a continuous improvement in the models' performance indicators, suggesting the merits of incremental learning. The proposed methodology improves predicted performance by continuously incorporating new data into the models, enabling for the capture of dynamic behavioural patterns among students. This is especially important considering the flexibility of student behaviours, which may be impacted by academic, social, or environmental circumstances. 

The emergence and re-emergence of Group 4 is especially interesting, given that only three groups were initially identified. It indicates the possibility of an evolution in the behavioural patterns of the students or the emergence of a distinctive group of behaviours that were not observed during the initial weeks. This observation underscores the dynamic nature of student behaviours and the importance of the grouping approach for spotting these changes. This variability may help understand the way in which student behaviours change as a result of different stages throughout an academic term. This emphasizes the importance of continuous monitoring to capture the complete range of behavioural patterns, which could be missed in a static approach. The consistent underperformance of Group 4 highlights concerns about the complexity of behaviours within this group, suggesting a possible hurdle for models in identifying patterns linked to loneliness. Furthermore, the varying levels of performance of models in Group 3 indicate the need for a range of algorithms that can effectively capture unique behavioural characteristics. These findings highlight the significance of selecting and training models that include the heterogeneity among behavioural groups. These models' intricate and nuanced performance underscores the complex nature of loneliness, emphasizing the need for interventions tuned to these distinct features. 

This study presents a number of advancements in the field of student mental health research by introducing an innovative approach for loneliness detection that employs a dynamic framework that continuously monitors student behaviour patterns, facilitating early identification of loneliness and enabling timely intervention. The integration of incremental clustering, incremental classification, and multi-model voting plays a crucial role in enhancing the detection of loneliness among college students. Incremental clustering efficiently tracks and adapts to the evolving behavioural patterns of students, ensuring the identification of students' behavioural groups process remains tuned to real-time changes. Incremental classification contributes by continuously updating and refining prediction models with incoming data, maintaining their effectiveness in a dynamic student environment. Multi-model voting combines the strengths of diverse models, yielding more balanced and accurate loneliness predictions by considering a wider range of behavioural indicators and reducing reliance on any single model's viewpoint. These methodologies collectively ensure a more nuanced, responsive, and accurate approach to loneliness detection in the college student population.

This work has important practical implications for students' health and society's overall well-being. It leverages a non-intrusive approach to gather sensing data to detect and deal with loneliness at an early stage. By using passive sensing data, this method may optimize university health services by detecting high-risk students, hence possibly enhancing students' overall well-being. The study's approach extends beyond the academic domain, with possible applications in many demographics, such as older people and organizations, to strengthen mental health initiatives. While novel, this study contains limitations that need careful interpretation and offer future research directions. The incremental learning approaches were evaluated on a 10-week student sample, emphasizing the need for validation on longer-term, diverse datasets that show reliability across different demographics. Relying on a single dataset risks not capturing the whole range of behavioural diversity, which limits the detection models' scalability. Furthermore, self-reported loneliness surveys, which are susceptible to recall bias and subjective interpretation, may not adequately reflect the complexities of loneliness experiences.

\subsection{Conclusion}
This study is a further step forward in comprehensively detecting loneliness, here among students. We have used passive sensing data combined with an innovative method integrating incremental clustering, classification, and multi-model voting. This approach effectively captures and adapts to changing student behaviours, offering a nuanced understanding of loneliness. Our findings show the superior performance of group-based models over generic models for this application, emphasizing the need for more personalized mental health interventions. However, the study's focus on a specific student demographic and a limited duration highlights the necessity for a broader and longer-term study to enhance its applicability. Future research should address the challenges of model bias, dynamic behavioural patterns, ethical considerations in passive data usage, and the application to other and also large participant populations.

%
%
%
%

\end{document}